\documentclass{sig-alternate}
\usepackage{graphicx}
\usepackage{subfigure}
\usepackage{url}

\begin{document}

\title{From Linked Data to Relevant Data -- Time is the Essence}

\numberofauthors{3}

\author{
\alignauthor
Markus Kirchberg\\
       \affaddr{Services Platform Lab}\\
       \affaddr{Hewlett-Packard Laboratories}\\
       \affaddr{Fusionopolis, Singapore}\\
       \email{Markus.Kirchberg@hp.com}
\alignauthor
Ryan K L Ko\\
       \affaddr{Services Platform Lab}\\
       \affaddr{Hewlett-Packard Laboratories}\\
       \affaddr{Fusionopolis, Singapore}\\
       \email{Ryan.Ko@hp.com}
\alignauthor
Bu Sung Lee\\
       \affaddr{Services Platform Lab}\\
       \affaddr{Hewlett-Packard Laboratories}\\
       \affaddr{Fusionopolis, Singapore}\\
       \email{Francis.Lee@hp.com}
}
\date{08 February 2011}

\maketitle
\begin{abstract}
 The Semantic Web initiative puts emphasis not primarily on putting data on the Web, but rather on creating links in a way that both humans and machines can explore the Web of data. When such users access the Web, they leave a trail as Web servers maintain a history of requests. Web usage mining approaches have been studied since the beginning of the Web given the log's huge potential for purposes such as resource annotation, personalization, forecasting etc. However, the impact of any such efforts has not really gone beyond generating statistics detailing who, when, and how Web pages maintained by a Web server were visited.

In this paper, we propose a different approach of how to mine Web server logs by combining link analysis and usage analysis for events based on time-windowed views over usage logs. We detail our observations and argue that the sum of all Web Travel Footprints' data on the visitors access paths and the linkage of the resources within the site at a particular time window gives sufficient insights at what constitutes relevance. Our findings are substantiated by presenting evidence in the form of exploratory graphics. 
\end{abstract}

%
%
%
\section{Introduction} \label{sec:intro}

The World Wide Web (WWW) has seen several major evolutions over the past 2 decades; coining terms such as the traditional Web, Web 2.0, and Web 3.0. While the WWW was originally designed as a distributed information space to support human navigation through linked documents, machine-based access has been hindered by the type of mark-up used within those documents. The Semantic Web (part of Web 3.0) is the most prominent current effort to address this shortcoming. It aims to bring the WWW to a state in which all its content can also be interpreted by machines. The emphasis is not primarily on putting data on the Web, but rather on creating links in a way that both humans and machines can explore the Web of data. Accordingly, the term Linked Data was coined \cite{Bizer.C:09:LD-SS}. Linked Data builds upon standard Web technologies, e.g., HTTP and URIs, but extends them to share information in a way that data from different sources can be connected and queried.

When users access the Web, they leave a trail as Web servers typically maintain a history of page requests they receive. Given the existence of resource request/hit histories, there have been numerous efforts in Web usage mining, i.e., trying to determine what users are interested in. This is to no surprise as there is a huge potential in evaluating usage histories for purposes such as annotation of Web resources, personalization of search or recommendation tools, forecasting of future page requests or usage patterns, and so on. However, the impact of any such efforts has not really gone beyond generating statistics detailing who, when, and how Web pages maintained by a Web server were visited.

In this paper, we propose a different approach of how to mine Web server logs by combining link and usage analysis for events based on time-windowed views over usage logs. We detail our observations and argue that this leads us to a notion of time-windowed relevance. In this context, the term event can be understood as a situation that creates a need in a human/machine to search or browse for related information which, in turn, triggers a visit to a Web resource that is associated with topics and keywords via the Web 3.0.

\subsection{Related Work} \label{sec:relatedWork}

It should be acknowledged that there exists a wealth of literature that is related to the work discussed in this paper. Given the scope and context of our effort, we will only briefly refer to the works that we considered most relevant to us.

Wang et.al. \cite{Wang.J:02:Rurtt} proposed a notion of relevance (w.r.t. a given topic) in the context of link analysis on Web logs. Relevance is obtained by considering the users' expertise and the importance of Web pages within a unified framework. Roughly speaking, a user's expertise correlates to the quantity and quality of Web pages visited, while the importance of a Web page is determined by citation of other pages and the frequency of visit by users. Thus, importance of Web pages and expertise of users influence each other.

Dupret et.al. \cite{Dupre.G:10:mteid} studied ranking/relevance not by counting document hits themselves, but rather by concentrating on the user behaviour after accessing a document.

Delbru et.al. \cite{Delbr.R:10:HLAfR} introduced a hierarchical link analysis approach for Linked Data. A two-layer model for ranking Web data is proposed. While this work does not consider Web usage logs, it has influenced our thinking of how to extend our approach to multiple data sets.

M{\"oller} et.al. \cite{Molle.K:10:LfLOD} presented their findings made during the analysis of four Linked Data usage log data sets, which include parts of the two logs used for our work.

\section{Motivation and Contribution} \label{sec:motivation}

The area of data analytics has seen a huge increase in interest driven by the emergence of the cloud computing and map-reduce data processing paradigms as well as the maturing and increased acceptance of semantic technologies. Having capabilities to process huge amounts of logged data, which are generated not only on the Web but by almost all systems and services in-use nowadays, will constitute a key distinguishing factor w.r.t. personalization, quality of service and competitiveness. Our current focus is on extracting meaningful information, such as usage patterns or relevance indicators, and relate it back to the Web and/or the users of the Web. While there have been a variety of approaches proposed that looked at page ranking, link analysis and usage log analysis, these approaches need to be revisited and/or rethought to keep up with the advances in technology. In this paper, we present initial evidence based on USEWOD 2011 Data Challenge data sets that usage logs can indeed lead to a notion of relevance.

Past approaches have typically considered time as an orthogonal factor. For instance, page ranking approaches (a form of link analysis) assign a numerical weighting to each element of an interlinked set of Web pages with the purpose of determining its relative importance within the set. While such rankings consider the importance of each page that casts a vote, giving some pages greater value over others, time is not primary -- often not even a secondary -- factor. This constitutes a limitation of the potential impact of the measured importance, in particular, if it is related to real-world events, topics or keywords that are not consistently interpreted over time, information that has strong temporal properties, and so on.

We perceive that it is essential not only to consider the interlinking of weighted Web resources, but also whether humans/machines make use of such links (i.e., putting less emphasis on the mere existence of a link or a link structure), how they utilise them (e.g., browsing depth within a data set, common browsing patterns within a single or across various data sets), and how the usage changes over time.

In this paper, we discuss the rationale behind our proposed approach, present evidence (using the USEWOD 2011 Data Challenge data set) that time is indeed a key factor that has to be considered, and detail initial findings, propositions and ways to move forwards.

\section{The USEWOD 2011 Data Set} \label{sec:usewod2011}

Our current research and propositions are based on the USEWOD 2011 Data Challenge data set\cite{Beren.B:11:U-1iw}. This data set is comprised of usage logs for two sites (i.e., Semantic Web Dog Food and DBpedia) that form a part of the Web of data. Usage logs conform to the Apache Combined Log Format (CLF), with the following modifications:

\begin{itemize}

 \item (anonymisation) All IP address fields have been replaced with `0.0.0.0';

 \item (added support for location-based analyses) A new field with the country code (using the GeoLite Country API from MaxMind) of the original IP address has been appended to each log entry; and

 \item (added support for distinguishing requestors) A hash of the original IP has been appended to each log entry.

\end{itemize}

\subsection{Evaluated Data Sets}

For our analysis, we had access to Web server usage logs from the DBpedia Linked Data site and the Semantic Web Dog Food (SWDF) Linked Data site. In addition, a prior analysis \cite{Molle.K:10:LfLOD} that included a subset of the USEWOD 2011 Data Set has provided initial clues of what can easily be extracted from the usage logs and, most importantly, how to distinguish conventional accesses from semantic accesses\footnote{A semantic access is reflected in the log files by two entries: 1) a request for the plain resource (URI), which is answered with the HTTP 303 redirection code; and 2) a request for the corresponding resource representation, which is answered with the HTTP 200 success code.}.

Before we discuss initial findings, let us briefly introduce the two Linked Data sites that form part of our analysis together with their core properties (depicted in Table \ref{tab:dataSets}) as they can be observed from the usage logs.

\begin{table*}
 \centering
 \begin{tabular}{|r||r|r|r|r|r|r|r|}
  \hline
           & Size of  & \# Accessed &         &              & \# Success-  & \# Semantic & \# SPARQL\\
  Data Set & Logs     & Resources   & \# Days & \# Hits      & ful Hits     & Requests    & Requests\\
  \hline
  \hline
  SWDF     & $2.02$GB & $40,322$    & $720$   & $8,092,552$  & $7,098,705$  & $384,163$   & $1,557,893$\\
  \hline
  DBpedia  & $6.96$GB & $5,976,545$ & $23$    & $19,770,157$ & $18,862,220$ & $55,640$    & $4,500,604$\\
  \hline
 \end{tabular}
 \caption{Core Properties of the USEWOD 2011 Data Challenge Usage Logs.}
 \label{tab:dataSets}
\end{table*}

\begin{figure*}[th]
 \centering
 \subfigure[Hits Overall]{
   \includegraphics[width=.3\textwidth]{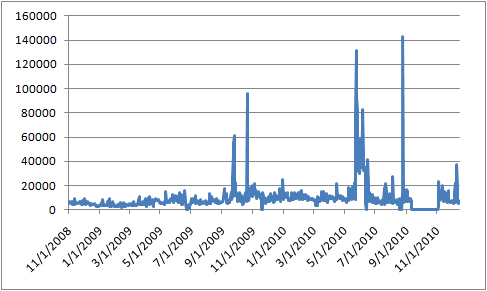}
   \label{fig:swdfHitsWithBots}
 }
 \subfigure[Hits and Trend w/o Bots \& Spiders]{
   \includegraphics[width=.3\textwidth]{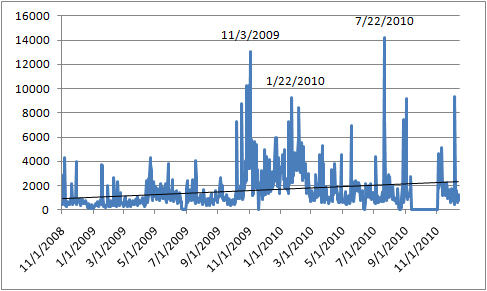}
   \label{fig:swdfHitsNoBotsTrend}
 }
 \subfigure[Normalised Hit Distribution$^3$]{
   \includegraphics[width=.3\textwidth]{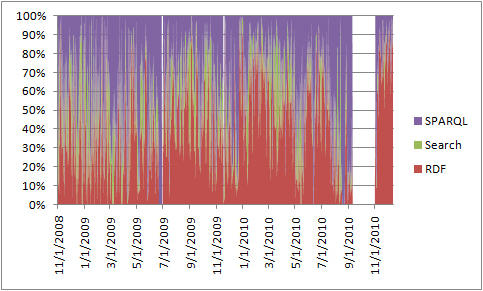}
   \label{fig:swdfHitsByNonHTMLAccessTypeNoBots}
 }
 \caption{Hit Statistics for SWDF}
 \label{fig:swdfHits}
\end{figure*}

\subsubsection{Semantic Web Dog Food}

The SWDF Linked Data site contains information about publications, people and organisations associated with the Web and Semantic Web domains. This data set covers some of the domains' main conferences and workshops, including WWW, ISWC and ESWC. The provided usage log files cover just over two years from Nov 01, 2008 to Dec 14, 2010 -- however, logs for about 50 days are missing. As detailed in Table \ref{tab:dataSets}, the logs contain entries for $720$ days and comprise just over $8$ million hits (out of which about $7$ million hits have a status code $2xx$ signalling success or $3xx$ signalling redirection) over about $40,000$ existing resources\footnote{For this paper, an existing resource (or from here on simply resource) is a Web resource that has had at least one hit with HTTP status code of type success or redirect. For our analysis, we have removed all hits with HTTP status code $4xx$ or $5xx$, which signal an error with the request/hit or the resource, as a first data cleaning step.}.

\subsubsection{DBpedia}

The DBpedia Linked Data site is a twin of Wikipedia and constitutes one of the focal points of the Web of data. DBpedia contains structured information from Wikipedia with the aim to enable automated querying against Wikipedia content as well as to serve as a central Linked Data node hosting commonly used terms, terminologies and concepts (to assist with the integration of knowledge hosted in different Linked Data sites) in the form of URI and RDF descriptions \cite{Bizer.C:09:LD-SS}. The DBpedia usage logs contain entries for $23$ days between Jul 01, 2009 and Feb 01, 2010. Compared to SWDF, DBpedia's usage logs contain a significantly larger number of hits (i.e., about $20$ million), successful hits (i.e., about $19$ million) and (existing) resources (i.e., about $6$ million).

\subsection{Evaluation Framework}

Given the sheer size of the logs and the complexity of the usage data, we decided to convert the plain-text usage logs into a more queryable representation and apply various steps of data cleaning and data enrichment. Our corresponding process can be summarised as follows:

\begin{enumerate}

 \item Log entries have been evaluated for their consistency and, if deemed suitable, broken up into their parts as contained in the extended CLF format. During this step, we have also removed all hits with HTTP status code $4xx$ or $5xx$ as we were not interested in determining common request errors, broken links etc.\\
  Initial observations were that the SWDF data set is very clean and conform to the CLF format (with only a single hit being discarded; due to a likely error in the data anonymisation process), while the DBpedia usage log contains a few thousand entries that are either not UTF-8 decodable or contain referrer/resource references with unencoded \texttt{"} characters breaking CLF formatting.
  
 \item Each log entry that was deemed clean and corresponded to a successful or redirection hit was then processed and entered into a PostgreSQL database specifically designed for our efforts.

 \item Once both usage log sets had been imported into the database, post-processing routines then identified:

  \begin{itemize}

   \item URIs and matching HTML/RDF representations;

   \item Bots, spiders, crawlers etc. -- based on information specified in the log's user agent field, access to a Linked Data node's \texttt{robots.txt} file and high frequency accesses (hosts that issued thousands of hits within a very short span of time); and

   \item Access types -- Plain/HTML versus Semantic versus Search (SWDF only) versus SPARQL.

  \end{itemize}

 \item Basic analysis of usage log data (in Section \ref{sec:usewod2011}).

 \item Relevance-driven usage log analysis (in Section \ref{sec:relevance}).

\end{enumerate}

\subsection{Basic Evaluation Results}

\begin{figure*}[t]
 \centering
 \subfigure[DBpedia: Normalised Hit Ratio]{
   \includegraphics[width=.3\textwidth]{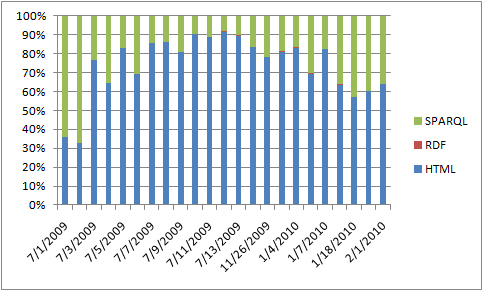}
   \label{fig:dbpediaHitsByAccessTypeNoBots}
 }
 \subfigure[DBpedia: Hits by Access Type]{
   \includegraphics[width=.3\textwidth]{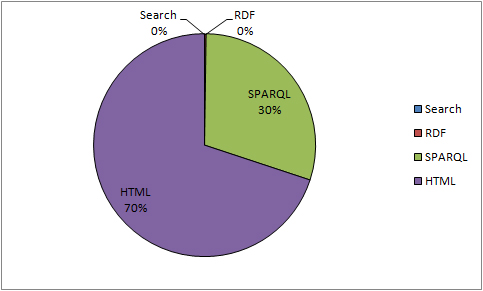}
   \label{fig:dbpediaAccessTypesNoBots}
 }
 \subfigure[SWDF: Hits by Access Type]{
   \includegraphics[width=.3\textwidth]{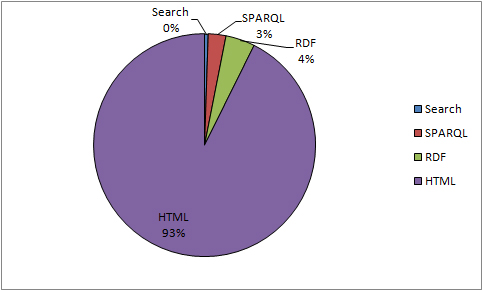}
   \label{fig:swdfAccessTypesNoBots}
 }
 \caption{Hit Statistics for DBpedia and SWDF (without considering Bots \& Spiders)}
 \label{fig:dbpediaHitsCountry}
\end{figure*}

Our initial efforts aimed at gaining an understanding of whether any straight-forward insights can be obtained from the available data sets. For this, we generated various hit statistics and looked at whether semantic accesses or SPARQL queries are a common occurrence in the Web of data.

M\"{o}ller et.al. \cite{Molle.K:10:LfLOD} already outlined initial findings for a subset of the SWDF and DBpedia data sets. During that analysis it was observed that `[...] \textit{conventional traffic on the Dog Food site has been increasing steadily since its inception in July 2008, while demand for RDF data has been more or less static on a comparably low level. In fact, true semantic requests are even lower than requests for RDF representations, which indicates that most agents requesting semantic data blindly follow links, without actually knowing what to do with the received RDF data.\/} [...]' \cite[page 4]{Molle.K:10:LfLOD}. We could confirm such observations across both data sets as depicted in Figures \ref{fig:swdfHitsByNonHTMLAccessTypeNoBots}, \ref{fig:swdfAccessTypesNoBots}, \ref{fig:dbpediaHitsByAccessTypeNoBots}, and \ref{fig:dbpediaAccessTypesNoBots}. Notably, the SWDF data set has a negligible number of non-Plain/HTML accesses\footnote{If we would add Plain/HTML accesses to Figure \ref{fig:swdfHitsByNonHTMLAccessTypeNoBots}, semantic accesses, SPARQL nor Search would be visible.} while SPARQL accesses are much more common with DBpedia. In fact, the ways both Linked Data nodes are intra-linked within themselves as well as how their respective requests pattern differ has implications on what one can infer from the corresponding usage logs. We briefly touch on this again later in section \ref{sec:relevance} when we outline properties that help to determine a notion of relevance.

Besides examining access types, M\"{o}ller et.al. also discussed a possible metric to determine relevance of a data set based on the level of influence that real-world events have on usage log statistics. However, it was concluded that `[...] \textit{In the case of the Dog Food dataset, the hypothesis is that requests for data from specific conferences would be noticeably higher around the time when the event took place.\/} [...] \textit{Contrary to our expectations, there are no significantly higher access rates around the time of the event.\/} [...]' \cite[page 4]{Molle.K:10:LfLOD}. Again, we were able to confirm those finding (as depicted in Figures \ref{fig:swdfHitsWithBots} and \ref{fig:swdfHitsNoBotsTrend}) and, more importantly, gain hints at how else to look at the data to be able to motivate a different notion of relevance. We will elaborate on our corresponding observations and discuss our approach in Section \ref{sec:relevance}.

Figures \ref{fig:swdfHitsWithBots} and \ref{fig:swdfHitsNoBotsTrend} highlight another interesting point -- the top hits in Figure \ref{fig:swdfHitsNoBotsTrend} excluding bots and spiders are $10\%$ of those found in the logs if bots and spiders are not removed. That is, being able to adequately filter bots, spiders, crawlers, high-frequency scripts etc. is vital to obtain a better insight into what users are really doing/looking for. However, it is not enough to already derive at a useful notion of relevance. For our subsequent analysis, we mainly focus on the usage logs data that are free from bots and spiders. Furthermore, we also consider access patterns (i.e., browsing histories by IP address/host hash derived from information about referring resources and accessed resources), which further clean the data from unwanted hits.

\section{Relevance} \label{sec:relevance}

In this section, we define the concept of \textit{relevance\/}, and demonstrate how the USEWOD 2011 data sets give us interesting clues and results which point to concepts of relevance of Web resources with time and events in reality.

First, let us clarify a few terminologies. We define \textit{two spaces\/} in which semantic data are communicated and stored. The first space is \textit{\bf Real Space\/}, where actual real-world events take place, at their unique time windows. A same semantic of an event, e.g. the National Day of a country, can take place yearly with the same objectives and content, but because of the different time windows, we are able to make sense of it, and understand its temporal and situational context/meaning. A 2011 National Day will be a different event from a 2010 National Day, and the distinguishing factor (even if all things are constant) will be the time period. The second space, we are concerned about, is \textit{\bf Web Space\/}, otherwise known as cyberspace. In the Web Space, linked data exist and encapsulate descriptions of the actual events in Real Space in the form of linked data (i.e., keywords, topics, Web pages). The current focus of the Semantic Web research is on inferring the meaning of the data from the data or by describing how they link to each other, but we argue that, without a time window, it is more difficult to give `Meaning' or semantics to a set of keywords or topics or Web data describing a Real Space event.

Hence, in this paper, we propose the time windowed concept of studying representations of events in Real Space recorded as linked data in Web Space. We notice that when we apply time windows (e.g., one week duration before a meaningful event, during the event etc.), we are able to observe a trend through exploratory graphics -- the relationship and meaning between traffic and resources of linked data actually stand out, and most importantly, change meaningfully over time windows. The strength of such a relationship between time window, traffic and the linked data resource is what we call the concept of Relevance.

\subsection{Approach}

So how do we represent Relevance? \ We can get pointed clues from our definition in Section \ref{sec:intro} earlier -- `An event can be understood as a situation that creates a need in a human/machine to search or browse for related information which, in turn, triggers a visit to a Web resource that is associated with topics and keywords via the Web 3.0.'

This means that if we can capture or visualise the whole knowledge seeking path from the point of interest of human users in an event in the Real Space, to the exploration for knowledge of event-related linked data within the Web Space, and eventually to the generation of surfing/access patterns of the Web user when looking for the knowledge of a topic (i.e., recorded as Web logs), we are able to infer how a Real Space event has relevance to certain topics in the linked data, and even how strong are the links between different linked data. From such understanding, we are able to quantify the strength, influence, and the linkages between topics, just from usage data.

This paper presents evidence of such a chain of inference in the form of exploratory graphics (i.e., our proposed Kandinsky Graphs in Section \ref{sec:kandinskyGraphs}). It is perhaps important to note the difference between exploratory graphics and presentation graphics. Exploratory graphics are visualisations that give important clues and observations of patterns and consistent trends, which eventually can be used to prove the existence or understanding of a certain phenomenon, and then model them as mathematics, algorithms or other formalisms that can reproduce such trends. Presentation graphics, on the other hand, are representations of hard facts, of phenomena already known to the public.

\subsubsection{Measuring Relevance} \label{sec:measuringRelevance}

Web logs give us the traffic data of who visited the site at what time, and where the access came from. But they do not give information about how relevant and closely-linked are each resource. However, if we view the traffic log data from another perspective, we are able to generate two important metrics useful for the calculation of relevance:

\begin{enumerate}

 \item \texttt{Fan} $=$ Number of resources this resource is currently linked to. 

 \item \texttt{Weight} $=$ Summation of all incoming and outgoing access traffic (irregardless of human or machine access).

\end{enumerate}

Whereby for fan and weight the following hold:

\begin{itemize}

 \item Strength of relevance of a resource $=$ A set of (1) the cumulative weights of all linkages to other resources, and (2) the magnitude of the Fan of the resource, within a time window.

 \item Most relevant resource to an event in a time window $=$ The resource with the largest strength of relevance within a Web-site, within a time window.

\end{itemize}

\subsubsection{Web Travel Footprint (WTF) of an IP Address}

\begin{figure*}[t]
 \centering
 \includegraphics[width=\textwidth]{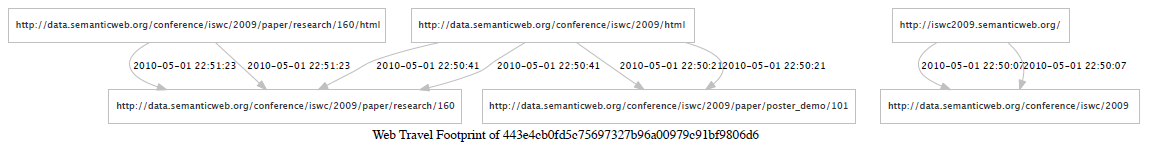}
 \caption{Web Travel Footprint of a single IP Address During WWW 2010 Time-window.}
 \label{fig:wtf}
\end{figure*}

Although we are able to retrieve some `identity' (i.e., IP address) of the visitors from log data, we recognise that a single IP address does not necessarily translate into a single user or machine. Nevertheless, it is an important representation of the location/source of the person or object that started a path of curiosity investigating data (in Web Space) on a certain event (in Real Space). To get meaningful paths for analysis of relevance, the Web-site has to be an active Web-site (i.e., with real human/semantic traffic), and must have a purpose (e.g., news site or discussion forums). The importance of an active and purposeful Web-site was observed in the USEWOD 2011 Data Challenge data set, via the contrast in results for DBpedia (a site with fewer meaningful footprints), and the SWDF Web-site (a site which is heavily accessed by users with a related interest).

A user's quest for knowledge during a visit to the pages of a site leaves behind a Web Travel Footprint (WTF) for his/her IP address, as shown in the example in Figure \ref{fig:wtf}. This is likened to a network of roads on a map, with the footprint being the trail taken by the user. Each WTF represents cumulated footprints left by each IP address, and represent three relevance characteristics distilled from two variables found from the logs (illustrated in Figure \ref{fig:kgCharacteristics}):

\begin{enumerate}

 \item From linking the `referrer' to `resource requested':

  \begin{enumerate}

   \item \textbf{Fan} -- Linkages between a data resource and other data resources.\\
    When we link the referrer data to the directory path of the Web resource requested (e.g., via GET), we can chart out the fan of one resource to other resources. By viewing the linkages of resources from usage logs in a time window, we are able to see how users' surfing footprints can show the interlinkages between the different resources. For example, at a page about a Semantic Web researcher's profile, a user can click on lots of links, e.g., from his/her institution information to even a link to the email of the administrator of the site. Users interested in the background of the researcher will click on the institution's link more, while they may even ignore the email link at the bottom of the page. In this way, the usage logs logging these accesses have told us the semantic linkages users perceived between Institution to the Researcher Profile page. The same type of linkage between a page and its links can also be captured from the usage logs. The links form the fan, which shows the spread of influence of a resource, and only includes linked resources that have been accessed.

   \item \textbf{Depth}.\\
    When considering a sequence of multiple linkages between referrer and the resource requested, we can also see a path of how the IP address travels within the site, especially how `deep' the IP address source surfs into the Web-site. With a depth notion, we can measure how `curious' people from an IP address are on a certain set of resources. In another angle of looking at curiosity, we can see the strength of the travel within the site. By surfing continuously (link after link), it shows how a user wishes to find out more information about a certain topic. Longer paths indicate a higher relevancy of the linked pages/resources. When combined with links between data resources, the depth information can somewhat let us have an idea of how much interest a Web page has generated at that time. Such an interest could perhaps been triggered from an event within that time window.  

  \end{enumerate}

 \item From counting the number of times a link is accessed in logs within a time window:

  \begin{enumerate}

   \item \textbf{Weight} -- Number of times a path was accessed.\\
    With the concept of the fan and the depth, we are equipped to count the number of times the linkages occurred between the resources. From one IP address, it is interesting to see that it leaves behind footprints represented in cumulated weights from repeated travels in certain sets of paths of resources in a site over a time window. 

  \end{enumerate}

\end{enumerate}

For interpretation of relevancy of usage logs to real-life events, \textbf{all three} relevance characteristics must be considered together, and not in isolation. When considered alone, the fan or depth or weight characteristics do not tell much information. However, when combined and viewed together, we can see how these characteristics strengthen each other.

\subsubsection{Kandinsky Graphs -- Exploratory Graphs Representing Relevance of a Site's Web Resources} \label{sec:kandinskyGraphs}

Individually, WTF cannot tell the whole story of relevance of topics and resources for a site. When we consider the sum of all WTFs' data on the visitors (i.e., IP addresses) access paths and the linkage of the resources within the site at a particular time window, we can form summary graphs, named as Kandinsky graphs\footnote{We name such graphs Kandinsky Graphs as they resemble the painting style of the abstract artist Wassily Kandinsky.}.

\begin{figure*}[t]
 \centering
 \subfigure[Week Before Submission]{
   \includegraphics[width=.21\textwidth]{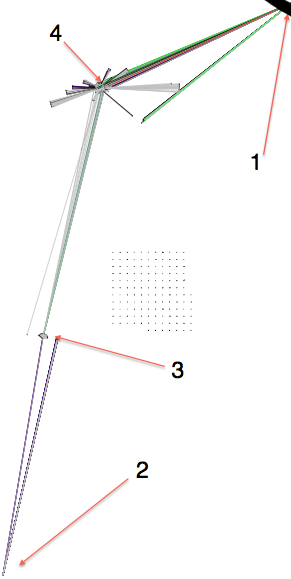}
   \label{fig:kgWWW2010submission}
 }
 \subfigure[Week Before WWW 2010]{
   \includegraphics[width=.3\textwidth]{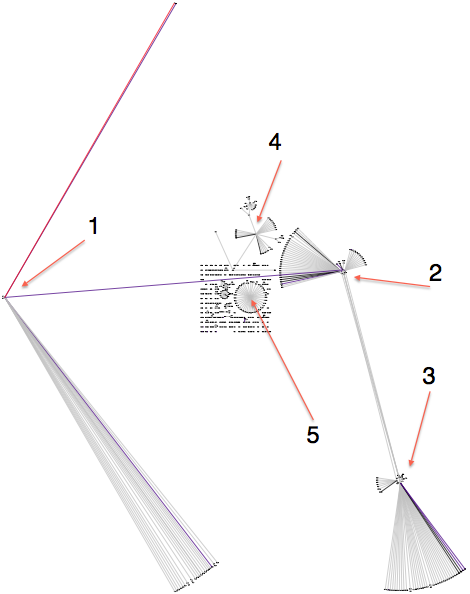}
   \label{fig:kgWWW2010wkBefore}
 }
 \subfigure[During WWW 2010]{
   \includegraphics[width=.3\textwidth]{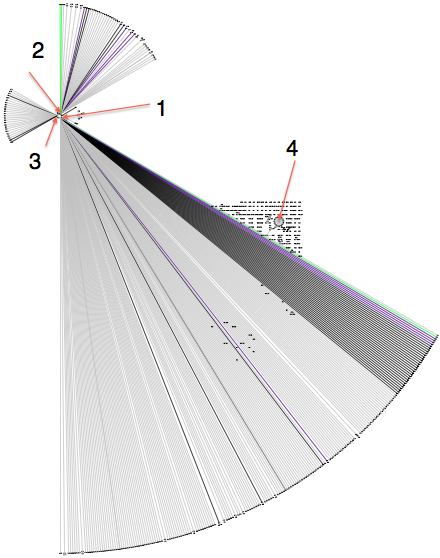}
   \label{fig:kgWWW2010during}
 }
 \caption{Kandinsky Graphs for WWW 2010}
 \label{fig:kandinskyGraphsWWW2010}
\end{figure*}

Kandinsky Graphs (KGs) are exploratory graphical summaries of (1) how deep users have travelled into and within the site, (2) how each resources are linked to each other and (3) which resources are highly relevant in the site -- at a given time window of a certain event, e.g., WWW 2010 conference (see Figures \ref{fig:kandinskyGraphsWWW2010} and \ref{fig:kgWWW2010wkAfter}). Kandinsky graphs are technically generated from GraphViz\footnote{\url{http://www.graphviz.org/}} dot files (created from our query results) as circo-layouts.

\begin{figure}[ht]
 \centering
 \includegraphics[width=\columnwidth]{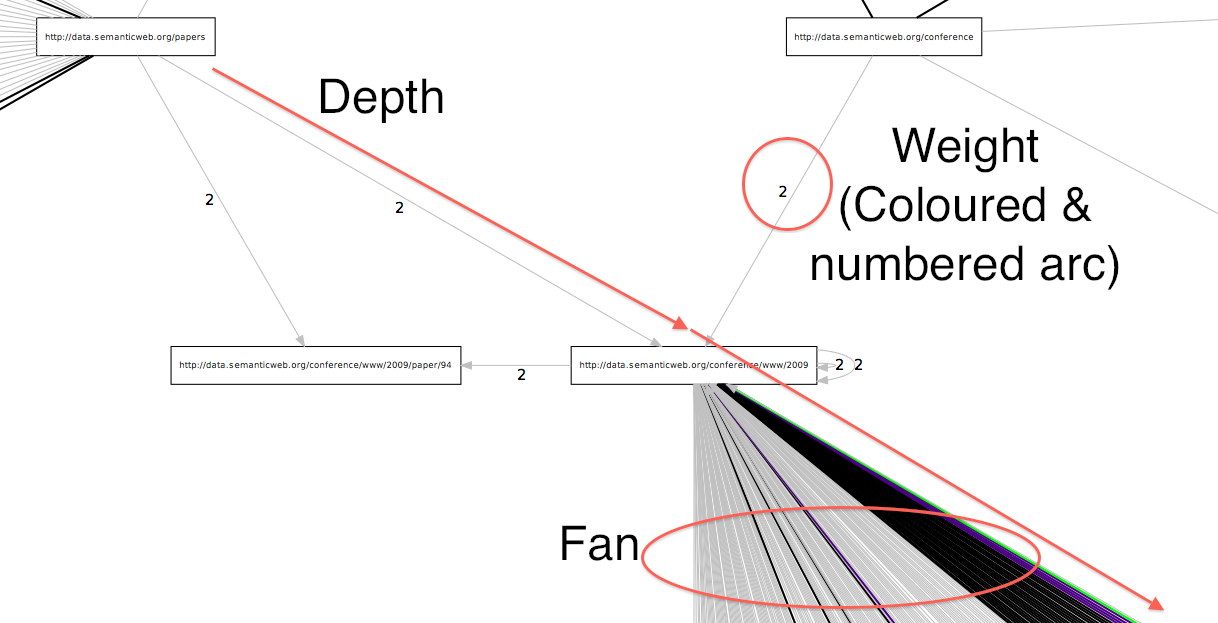}
 \caption{KG Characteristics (for Figure \ref{fig:kgWWW2010during}).}
 \label{fig:kgCharacteristics}
\end{figure}

Sites can be visited via any of its resources, and not necessarily be visited from the root index page. In the examples in Figure \ref{fig:kandinskyGraphsWWW2010}, the relevance of the resources in the defined time window can be visualised and observed via three previously mentioned metrics to observe (as per Figure \ref{fig:kgCharacteristics}): (1) the `fan' -- the number of links from a root node to different nodes; (2) the `weight' -- the number of occurrences as determined by the weight and colour coding of the edge between 2 nodes -- with red being the largest number of occurrences; and (3) the paths of access, and their depth (depicted by the number of resources linked by each single line). The colour coding of the weights of edges in the KG are as follows: gray (weight $\le$ 2),  black (2 $<$ weight $\le$ 4),  indigo (4 $<$ weight $\le$ 10), green (10 $<$ weight $\le$ 20) and red (weight $>$ 20).

It is now evident that from Kandinsky Graphs, the authors were able to prove important observations about which resources are the most relevant in a certain time period of a Real Space event. It also proves that usage logs can give semantic information by considering data boundaries defined by time windows. Besides observing the most influential and highly relevant resources in terms of fan and weight, we can even observe the \emph{`Scatter'} of resources and their closeness in relevance to the major topics of relevance. It is also interesting to see standalone clusters of linked resources, meaning that some resources are not even linked in terms of usage data (i.e., not relevant to each other)! 

\subsubsection{DIFF Kandinsky Graphs -- Showing the Difference Between Time Windows}

\begin{figure*}[t]
 \centering
 \subfigure[Week After WWW 2010]{
   \includegraphics[width=.3\textwidth]{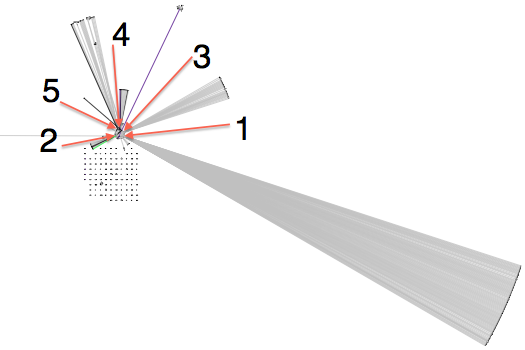}
   \label{fig:kgWWW2010wkAfter}
 }
 \subfigure[DIFF Before-During WWW 2010]{
   \includegraphics[width=.3\textwidth]{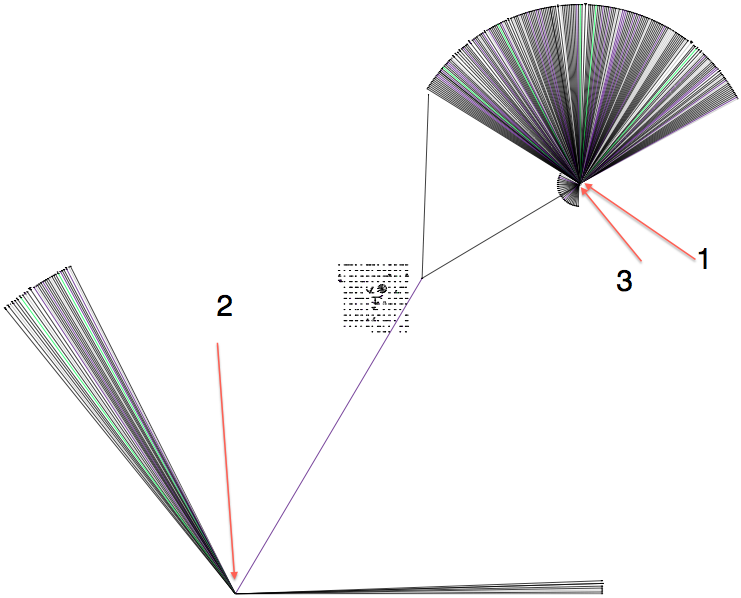}
   \label{fig:diffkgWWW2010beforeDuring}
 }
 \subfigure[DIFF During-After WWW 2010]{
   \includegraphics[width=.3\textwidth]{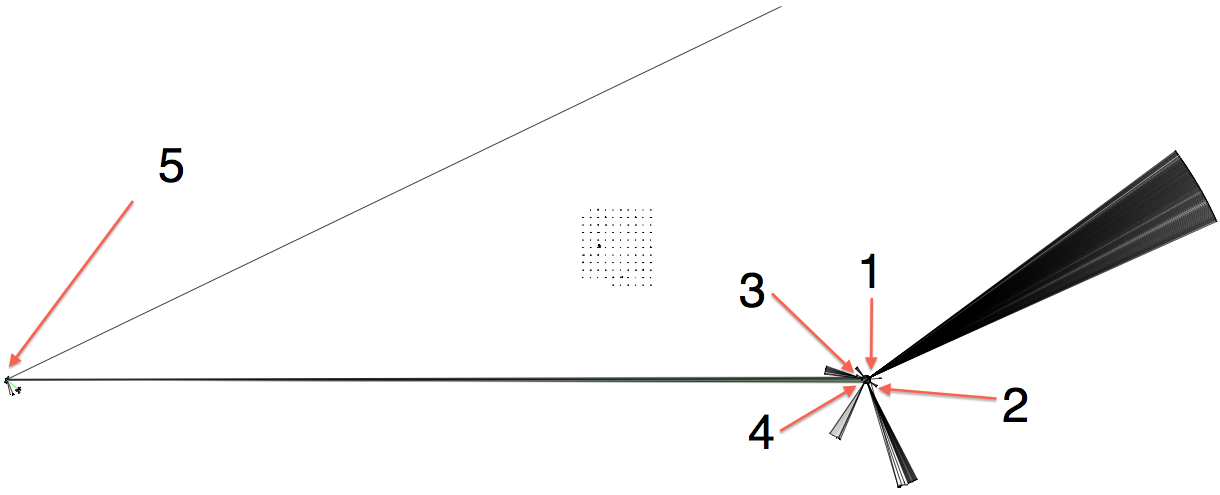}
   \label{fig:diffkgWWW2010duringAfter}
 }
 \caption{Remaining Kandinsky Graph and DIFF Kandinsky Graphs for WWW 2010}
 \label{fig:diffKandinskyGraphsWWW2010}
\end{figure*}

While the representation of relevance for each time window is captured in Kandinsky Graphs, we would also like to see the changes between time windows more obviously. This is the basis for the DIFF Kandinsky Graphs. With the DIFF Kandinsky Graphs, we aim to do the following:\\[1ex] $Relevance (TimeWindow_2) - Relevance (TimeWindow_1)$,\\[1ex] whereby weights are calculated using division (dropping the remainder values, i.e., removing edges from the graph that have seen a sharp drop between the old and the new time windows). That is, DIFF KGs emphasise on new hits and remove/heavily penalise edges with similar hits. For now, we are automating the calculations via a Python script on the PostgreSQL database.

With such information, we are able to generate Kandinsky Graphs for resources that are of new interest with each transition of time window (e.g., between the window before the event to the window during the event). The DIFF Kandinsky Graphs (see Figures \ref{fig:diffkgWWW2010beforeDuring} and \ref{fig:diffkgWWW2010duringAfter}) make the newly relevant resources stand out. The colour coding of the weights of edges in the DIFF KG are as follows: gray (weight $\le$ 1),  black (1 $<$ weight $\le$ 2),  indigo (2 $<$ weight $\le$ 5), green (5 $<$ weight $\le$ 10) and red (weight $>$ 10).

\newpage
\subsection{Case Studies}

\subsubsection{SWDF Data Case Study -- Relevance of SWDF Resources to the WWW 2010 Conference}

To further illustrate our point, let us study the Kandinsky Graphs of the WWW 2010 time windows w.r.t. the usage data of the SWDF Web-site.  WWW 2010, like any other academic conference, usually will have a few crucial time windows: (1) The week prior to the paper submission deadline; (2) The week before the WWW 2010 conference; (3) The actual WWW 2010 Conference; and (4) The week after the WWW 2010 conference. After the Kandinsky Graphs of the four time windows were generated (see Figures \ref{fig:kandinskyGraphsWWW2010} and \ref{fig:kgWWW2010wkAfter}), we were able to observe the change of rankings of the most relevant resources of the SWDF site for WWW 2010 windows in Table \ref{tab:swdfWWW2010CaseStudy}. The two DIFF Kandinsky graphs of the differences between the WWW 2010 conference time windows (before, during and after) depicted in Figures \ref{fig:diffkgWWW2010beforeDuring} and \ref{fig:diffkgWWW2010duringAfter} were also generated.  

\begin{table*}
 \centering
 \begin{tabular}{|l|c|c|c|c|l|} \hline
  Recurring Top Relevant Resources in the           & Week Before & Week Before & During   & Week After\\
  SWDF Web-site                                     & Submission  & WWW 2010    & WWW 2010 & WWW 2010\\ 
  \hline
  http://data.semanticweb.org/conference/www/2009/  & 2           & 2           & 1        & 3\\
  \hline
  http://data.semanticweb.org/conference/iswc/2009/ & 1           & 1           & 2        & 2\\
  \hline
  http://data.semanticweb.org/papers                & 3           & 3           & 3        & 4\\
  \hline
  http://data.semanticweb.org/person                &             & 4           & 4        & 5\\
  \hline
  http://data.semanticweb.org/organization          &             & 5           &          & \\
  \hline
  http://data.semanticweb.org/index.html            &             &             &          & 1\\
  \hline
 \end{tabular}
 \caption{Ranks of Relevance of Resources in SWDF in Relation to WWW 2010}
 \label{tab:swdfWWW2010CaseStudy}
\end{table*}

\paragraph{WWW 2010 submission (Oct 25 - 31, 2009)}
Refering to Figure \ref{fig:kgWWW2010submission}, we can observe that during the week before submission deadline, the ISWC 2009 page (Label 1) was the most relevant, followed by the WWW 2009 page (Label 2), then the main page linking to papers (publications) (Label 3), and in fourth place, the links to ESWC 2007 (Label 4). This tells us that there is a high interest in the real space in the publications of related conferences in the previous years. 

However, as ISWC 2009 was held between 25 October to 29 October 2009 (i.e. same time window within the WWW submission time window), it was also relevant to the SWDF Web-site.  This highlights the need for a relevance check to consider other overlapping time windows, and this will be investigated in our future work. 

\paragraph{Week before WWW 2010 (Apr 18 - 24, 2010)}
One week before the actual WWW 2010 conference (i.e., Figure \ref{fig:kgWWW2010wkBefore}), there is deeper interest in WWW 2009 (Label 2) information, related events such as ISWC 2009 (Label 1), and also in papers (Label 3), persons (Label 4), and organisations (Label 5) listed within SWDF. Perhaps some users are interested to check out the background of authors in accepted papers, or even to investigate possible related institutions for future collaborations. 

\paragraph{During WWW 2010 (Apr 25 - May 1, 2010)}
In Figure \ref{fig:kgWWW2010during}, we can see another change. This time, it was obvious that the WWW 2009 (Label 1) was the most relevant resource to the WWW 2010 conference in real space.  

\paragraph{Week after WWW 2010 (May 2 - 8, 2010)}
Figure \ref{fig:kgWWW2010wkAfter} shows an interesting phenomenon in the week after the WWW conference. Just after the conference, the most relevant resource became the main index page (Label 1) of the Web-site, and there were still sustained interest in papers and persons recorded in SWDF. Users' interest in the data related to the WWW event waned, and the traffic are no longer influenced by WWW 2010's occurrence.

\subsubsection{Observations from the DIFF Kandinsky Graphs}
We are interested in the differences between 1 week before the WWW 2010 conference and during the conference, and the difference between during the conference and 1 week after the WWW 2010 conference. From the DIFF Kandinsky Graphs in Figures \ref{fig:diffkgWWW2010beforeDuring} and \ref{fig:diffkgWWW2010duringAfter}, we were able to observe the following orders of relevance:

\paragraph{Difference between Time Windows: Before \& During}
\indent 1. \url{http://data.semanticweb.org/conference/www/2009}\\
\indent 2. \url{http://data.semanticweb.org/conference/iswc/2009}\\
\indent 3. \url{http://data.semanticweb.org/papers}\\

We could see that the difference between during and before the conference was in the WWW 2009 page and the papers. This mirror real-life events, and has further strengthened our arguments. 

\paragraph{Difference between Time Windows: During \& After}
\indent 1. \url{http://data.semanticweb.org/index.html}\\
\indent 2. \url{http://data.semanticweb.org/person}\\
\indent 3. \url{http://data.semanticweb.org/papers}\\
\indent 4. \url{http://data.semanticweb.org/conference/www/2009}\\
\indent 5. \url{http://data.semanticweb.org/conference/www/2010}\\

An interesting observation can be seen in the above order: The emergence of the WWW 2010 page right after the WWW 2010 conference. There was also a significant increase in interest in people and papers in the system.

\subsubsection{Other Case Studies}

From such observations, we decided to also study relevance of resources of the ISWC 2010 conference and the Haiti Earthquake via generating Kandinsky Graphs. However, due to space constraints, we are unable to include them in this paper, but have hosted our files on the following Web-site: \url{http://usewod2011.thekirchbergs.info/}.

From the ISWC 2010 conference's time windows, it was observed that there is indeed relevance between events in Real Space and time-windowed usage data in Web Space. The same changes of interest were clearly shown in the transitions across the time windows. It was found that there was also a high interest in the previous year's conference Web-site during the time windows before and during the ISWC 2010 conference. In the time window immediately after the conference, we also observed a sudden change of most relevant resources to a non-ISWC related resource.\\

After the successful tests and observations of relevance of resources with Real Space events in SWDF data, we decided to generate Kandinsky Graphs for the DBpedia Web-site during the time windows (before, during and after) the 2010 Haiti earthquake disaster.

The hypothesis was that there should be some relevance to the resources describing Haiti or earthquakes. However, this was proven wrong immediately via the Kandinsky Graphs. From the results, we could see that the top five (or even ten) most relevant resources were not anywhere remotely relevant to the topics relating to the Haiti Earthquake despite it containing Wikipedia's data. We further tested against other phenomenon (e.g., Michael Jackson's funeral in 2009) which occurred during the dates of the available DBpedia data sets, and generated time windowed Kandinsky graphs for them. A consistent observation of no relevance between Real Space events and DBpedia data was observed. 

On further analysis on the type of hits/requests made, we found that while DBpedia is large, it did not fulfil both the active and purposeful requirements mentioned earlier. SWDF, on the other hand, has a large pool of active users who were fairly consistent in interests (i.e., Semantic Web research), and was a site serving to provide information for Semantic Web research. Visitors to the site, minus the bots, have a purpose of seeking knowledge from SWDF.  

\subsection{Importance of Observations}

Our results and observations of relevance in active and purposeful Web-sites could only be achieved because of the fundamental linkage of time windows to the study of semantics in linked data. It represents a small but crucial step towards the identification of data relevant to real-life events from previously deemed contextless data such as Web logs. Such an observation will also greatly aid analytics research including sentiment analysis and quick identification of security-related events.

\section{Conclusions \& Future Work}
Based on the USEWOD 2011 Data Challenge data set, we have motivated a new approach to find a notion of relevance. We argue that the sum of all Web Travel Footprints' data on the visitors access paths and the linkage of the resources within the site at a particular time window gives sufficient insights at what constitutes relevance; important properties include: the fan, i.e, linkages between a data resource and other data resources; the depth of traversals; and the weight, i.e., number of times a path was accessed within a time windows. Our findings are substantiated by presenting evidence in the form of exploratory graphics. 

It must be acknowledge that our propositions are based on a small data set only and we need to further validate our findings via empirical studies using larger data sets from various domains. It is also necessary to refine the proposed approach in a way that one can differentiate between different events that overlap in time. 

Another set of problems comes into play when extending our notion of relevance to multiple data nodes. We foresee that a three layered approach is necessary to achieve this -- we need to understand the usage log, the composition and properties of the individual data nodes, and the way different data nodes are interlinked.

%
%
\small 
\bibliographystyle{abbrv}
\bibliography{Web}

\begin{thebibliography}{1}

\bibitem{Beren.B:11:U-1iw}
B.~Berendt, L.~Hollink, V.~Hollink, M.~Luczak-R\"{o}sch, K.~H. M\"{o}ller, and D.~Vallet.
\newblock Usewod2011 -- 1st int'l workshop on usage analysis and the web of data.
\newblock In {\em 20th Int'l World Wide Web Conference}, 2011.

\bibitem{Bizer.C:09:LD-SS}
C.~Bizer, T.~Heath, and T.~Berners-Lee.
\newblock Linked data - the story so far.
\newblock {\em Int'l Journal on Semantic Web andInformation Systems}, 5(3):1--22, 2009.

\bibitem{Delbr.R:10:HLAfR}
R.~Delbru, N.~Toupikov, M.~Catasta, G.~Tummarello, and S.~Decker.
\newblock Hierarchical link analysis for ranking web data.
\newblock In {\em 7th Extended Semantic Web Conference}, vol. 6089 of {\em LNCS}, pages 225--239. Springer, 2010.

\bibitem{Dupre.G:10:mteid}
G.~Dupret and C.~Liao.
\newblock A model to estimate intrinsic document relevance from the clickthrough logs of a web search engine.
\newblock In {\em 3rd Int'l Web Search and Data Mining Conference}, pages 181--190, ACM, 2010.

\bibitem{Molle.K:10:LfLOD}
K.~M\"{o}ller, M.~Hausenblas, R.~Cyganiak, S.~Handschuh, and G.~A. Grimnes.
\newblock Learning from linked open data usage: Patterns \& metrics.
\newblock In {\em Proceedings of the Web Science Conference (WebSci)}, 2010.

\bibitem{Wang.J:02:Rurtt}
J.~Wang, Z.~Chen, L.~Tao, W.-Y. Ma, and L.~Wenyin.
\newblock Ranking user's relevance to a topic through link analysis on web logs.
\newblock In {\em 4th Web Information and Data Management Workshop}, pages 49--54. ACM, 2002.

\end{thebibliography}

\end{document}